\documentclass[%
 reprint,
 superscriptaddress,
 amsmath,amssymb,
]{jfm}
\usepackage{amsmath}
\usepackage{amssymb}
\usepackage{graphicx}
\usepackage{epstopdf, epsfig}
\usepackage{graphicx}
\usepackage{dcolumn}
\usepackage{bm}
\usepackage{verbatim} 
\usepackage{microtype}
\usepackage[svgnames]{xcolor}
\usepackage{siunitx}
\usepackage{amsfonts}
\usepackage{blindtext}
\usepackage{hyperref}
\usepackage{mathptmx}
\usepackage{array} 
\usepackage{diagbox}
\usepackage{tabularx}

\DeclareMathOperator{\sinc}{sinc}
\renewcommand{\vec}[1]{\mathbf{#1}}
\newcommand{\trans}{\mathsf{T}}
\newcommand{\bu}{\vec{u}}
\newcommand{\bv}{\vec{v}}
\newcommand{\bw}{\vec{w}}
\newcommand{\br}{\vec{r}}
\newcommand{\bk}{\vec{k}}
\newcommand{\bx}{\vec{x}}
\newcommand{\bg}{\vec{g}}
\newcommand{\tbw}{\tilde{\bw}}
\newcommand{\kmin}{k_\text{min}}
\newcommand{\kmax}{k_\text{max}}

\hypersetup{colorlinks,breaklinks,
            urlcolor=Teal,
            linkcolor=MediumBlue,
            citecolor=NavyBlue,
            pdfauthor={Danyun He, Gautam Reddy, and Chris H. Rycroft}
            pdftitle={Energy-positive soaring using transient turbulent fluctuations}
}

\graphicspath{{./figs_orig/}}

\title{Energy-positive soaring using transient turbulent fluctuations}

\author{Danyun He\aff{1,2}
\corresp{\email{danyunhe@g.harvard.edu}},
Gautam Reddy\aff{3,4,5}
\and Chris H. Rycroft \aff{2,6}
\corresp{\email{chr@math.wisc.edu}}
}

\affiliation{\aff{1}John A.\@ Paulson School of Engineering and Applied Sciences, Harvard University, Cambridge, MA 02138, USA
\aff{2} Department of Mathematics, University of Wisconsin--Madison, Madison, WI 53706, USA

\aff{3} Physics \& Informatics Laboratories, NTT Research, Inc., Sunnyvale, CA 94085, USA
\aff{4} Center for Brain Science, Harvard University, Cambridge, MA 02138, USA
\aff{5} NSF-Simons Center for Mathematical \& Statistical Analysis of Biology, Harvard University, Cambridge, MA 02138, USA

\aff{6}Computational Research Division, Lawrence Berkeley Laboratory, Berkeley, CA 94720, USA}

\begin{document}

\maketitle

\begin{abstract}
Soaring birds gain energy from stable ascending currents or shear. However, it remains unclear whether energy loss due to drag can be overcome by extracting work from transient turbulent fluctuations. We designed numerical simulations of gliders navigating in a kinematic model that captures the spatio-temporal correlations of atmospheric turbulence. Energy extraction is enabled by an adaptive algorithm based on Monte Carlo tree search that dynamically filters acquired information about the flow to plan future paths. We show that net energy gain is feasible under realistic constraints. Glider paths reflect patterns of foraging, where exploration of the flow is interspersed with bouts of energy extraction through localized spirals.
\end{abstract}


\section{Introduction}
Soaring birds harvest energy by strategically gliding through atmospheric flows while extracting energy contained in the flows. For soaring birds like herring gulls and albatrosses, gliding consumes oxygen at a rate $\approx$30\% lower than flapping~\citep{baudinette1974energy, sakamoto2013heart}, which is crucial for making long-distance migration feasible within metabolic and aerodynamic constraints~\citep{tucker1972metabolism}.
The energetics of different forms of soaring can be described by a general expression for the non-dimensionalized rate of energy gained, $\dot{\varepsilon}$, by a glider (that generates no thrust) in a wind field:
\begin{equation}
    \dot{\varepsilon} = -c_D v^3 + w_z - \bm{v}.\dot{\bm{w}},
    \label{eq:energyrate}
\end{equation}
where gravity points in the negative $z$ direction, $c_D$ is the aerodynamic drag coefficient, $v = \|\bm{v}\|$ is the airspeed of the glider and $\bm{w}=(w_x,w_y,w_z)$ is the wind velocity~\citep{henningsson2011aerodynamics,harvey2021aerodynamic,taylor2016soaring}. (See section \ref{sec:glider} for details.) In the absence of wind, energy is continuously lost due to drag and the glider sinks at a constant rate. Thus, in order to compensate for drag, gliders should either actively localize at updrafts or align themselves anti-parallel to rapid gusts and wind shear along the glider's trajectory. These two mechanisms correspond to the second and third terms in Eq.~\eqref{eq:energyrate} respectively. Birds exploit both these mechanisms to gain energy, which correspond to two commonly observed modes of soaring known as thermal and dynamic soaring.

Thermal soaring relies on ascending currents (thermals) generated by convection in the atmospheric boundary layer~\citep{cone1962thermal,shannon2002american,woodward1956theory}. Thermals are dynamic flow structures that typically last a few minutes, providing updrafts that enable a bird to spiral up the boundary layer and forage for prey or glide to another thermal during migration~\citep{williams2018vultures}.
Adult vultures have more control on centering thermals than juveniles, suggesting that vultures improve their soaring skills through experience~\citep{harel2016adult}. Dynamic soaring allows for energy-neutral flight over oceans, where thermals are weak or absent~\citep{richardson2018flight, bousquet2017optimal,kempton2022optimization}. The predominant contribution to energy extraction during dynamic soaring is through the third term in Eq.~\eqref{eq:energyrate}~\citep{zhao2004optimal}, that is, by maintaining an appropriate heading while the bird manoeuvres through a stable shear layer generated behind ocean waves.

Nevertheless, both thermal and dynamic soaring rely on the formation of relatively stable convective plumes or wind shear. Atmospheric turbulent flows contain short-lived eddies of multiple time scales with velocity amplitudes comparable to a glider's typical sink rate, offering a potential continuous source of energy even in environments without stable convection or shear~\citep{mallon2016flight,patel2009extracting}. It is unclear whether energy can be extracted solely from the rapidly fluctuating wind fields that characterize such turbulent flows~\citep{reynolds2014wing,laurent2021turbulence}.
In order to harvest energy effectively, a glider should first acquire information about the flow, identify potential energy sources and select its path to maximize energy gain. However, acquired information degrades with time in an unsteady flow, which imposes strong constraints. In the extreme case of an uncorrelated flow, the value of information degrades immediately; consequently, $\dot{\varepsilon}$ is negative on average, $\langle \dot{\varepsilon} \rangle \simeq -c_Dv^3$. This is no longer true when the flow has correlations longer or comparable to the typical control timescale of the glider. What then is the relationship between the maximal energy rate, $\langle \dot{\varepsilon} \rangle_{\text{max}}$, the dynamic properties of the flow and the aerodynamic constraints on a glider? Even if energy-positive soaring is physically plausible, i.e., $\langle \dot{\varepsilon} \rangle_{\text{max}} > 0$, can a computational algorithm feasibly attain this limit?

Previous studies have examined navigational strategies for dynamic soaring in a static shear layer~\citep{bousquet2017optimal, kempton2022optimization} and thermal soaring in convective flows with~\citep{reddy2016learning, reddy2018glider} and without turbulence~\citep{wharington1998autonomous, allen2007guidance,chung2015learning}.  In the former case, a classical result from Rayleigh establishes the environmental conditions required to achieve energy-neutral flight~\citep{rayleigh1883soaring}. For thermal soaring, reinforcement learning methods have proved fruitful for identifying useful cues and effective navigational strategies in the face of turbulence~\citep{reddy2016learning, reddy2018glider}. These settings consider soaring in a convective flow, where the ascending branch is on the scale of hundreds of meters and lasts for minutes. Here, turbulence plays a disruptive role by introducing `noise' when identifying and localizing within relatively stable thermals~\citep{woodbury2014autonomous,akos2010thermal}. A recent proposal to extract energy from unsteady flows relies on a phenomenon known as fast-tracking exhibited by Stokesian inertial particles~\citep{bollt2021extract}. Online algorithms that learn the flow in real time to adaptively plan subsequent paths, and more generally, methods for active navigation in complex flows remain unexplored.
In this work, we address these questions using a kinematic model of three-dimensional homogeneous and isotropic turbulence. We consider a glider navigating within this flow that orients itself in response to its sensory history using a general-purpose decision-making algorithm that combines statistical inference and long-term planning. We show that energy-positive soaring is feasible and delineate the aerodynamic and flow parameters where this can be achieved.

\section{Model}
In this section we introduce the governing equation for a glider moving in a
wind field and the model for the turbulent wind field, comprised of
stochastically evolving Fourier modes. A number of aspects of our simulations
are computationally intensive, and our model is therefore implemented using a
custom C++ that uses OpenMP~\citep{dagum98} for multi-threading key
computations. Several implementation details are discussed in the appendices.
Tables~\ref{tab:param1} and \ref{tab:param2} summarize all of the parameters
used in the paper.


\subsection{Glider aerodynamics}\label{sec:glider}
\subsubsection{Governing equations for the glider}
\begin{figure}
  \begin{center}
    \includegraphics[width=0.4\textwidth]{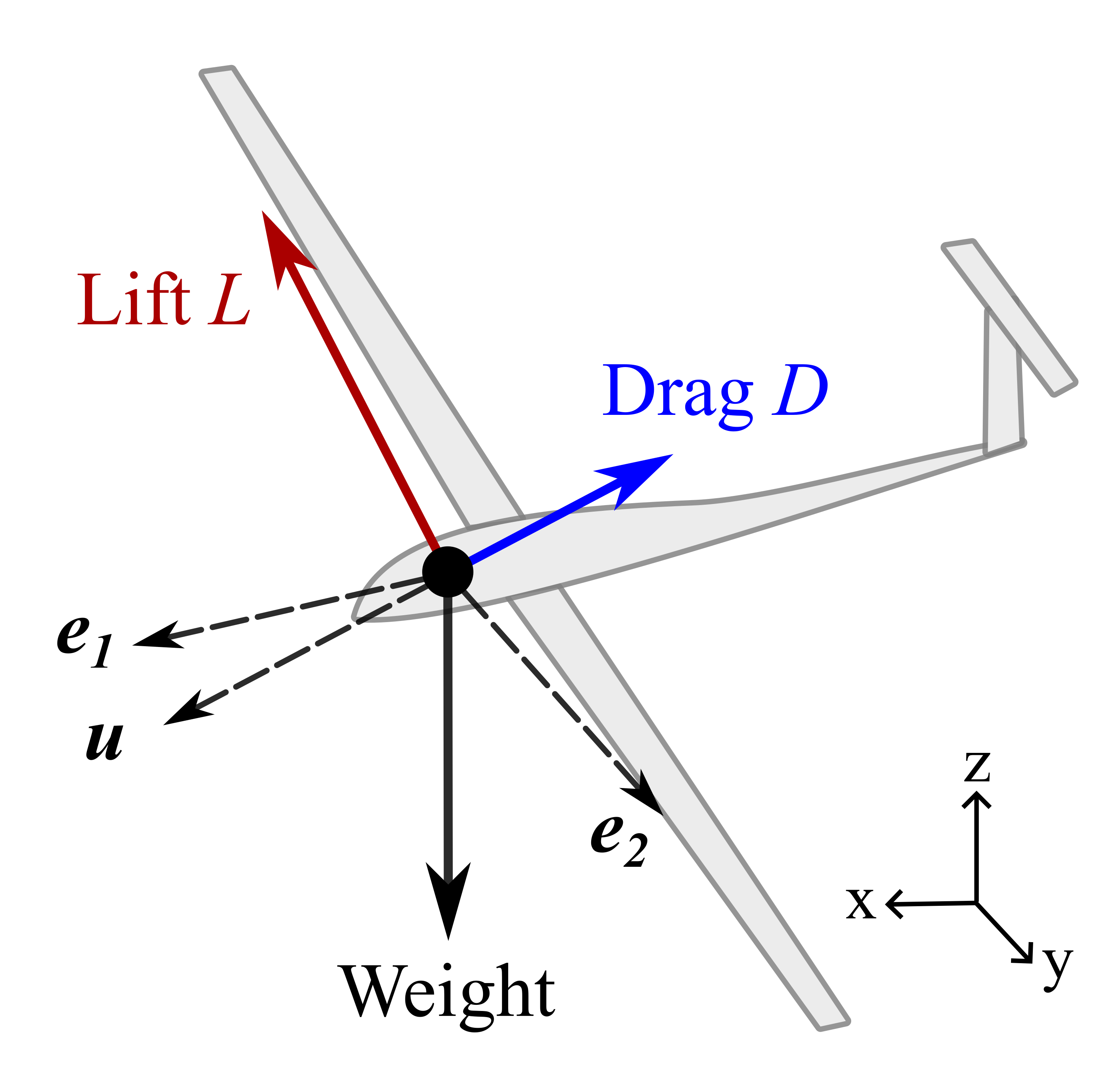}
  \end{center}
  \caption{Diagram of the glider, marking the directions of the three forces, lift $L$, drag $D$, and weight that it experiences. The global coordinate system is $(x,y,z)$, with the positive $z$ direction pointing vertically upwards. The $e_1$ vector points horizontally in the direction that the glider faces, although the glider direction may also have an additional $z$ component. The vector $\bu$ is the glider's velocity relative to the ground. The vector $e_2$ is horizontal, and is perpendicular to $e_1$.\label{fig:glider}}
\end{figure}

We simulate a glider moving in a global $(x,y,z)$ coordinate system where the $z$
axis points upward. The glider moves with time-dependent position $\br(t)$ and
velocity $\bu(t)$ relative to the ground, experiencing a local wind velocity
$\bw(t)$. The glider's motion relative to the wind is therefore given by $\bv =
\bu - \bw$, and the airspeed is given by $v=\|\bv\|$. As shown in
Fig.~\ref{fig:glider}, there are three forces acting on the glider: lift, drag,
and weight. Lift is a component of the aerodynamic force, and is directed
perpendicular to the flight direction. The magnitude of the lift force is given by
\begin{equation}
  L=\frac{1}{2}c_L\rho S v^2,
\end{equation}
where $\rho$ is the density of air, $S$ is the surface area of the wing, $v$ is the airspeed and $c_L$ is the dimensionless lift coefficient which has complex dependencies on wing shape. The drag is another component of the aerodynamic force, directed opposite to
the flight direction. It scales similarly to lift, and has magnitude
\begin{equation}
  D=\frac{1}{2}c_D \rho S v^2
\end{equation}
where $c_D$ is a dimensionless drag coefficient. Figure~\ref{fig:glider} also
defines two orthogonal vectors $(e_1,e_2)$ in the horizontal plane, with $e_1$
pointing in the same direction in the $xy$-plane as the glider body. Note
however that the glider may also be tilted, so that its direction vector has
an additional vertical component.

\begin{figure}
  \begin{center}
    \includegraphics[width=\textwidth]{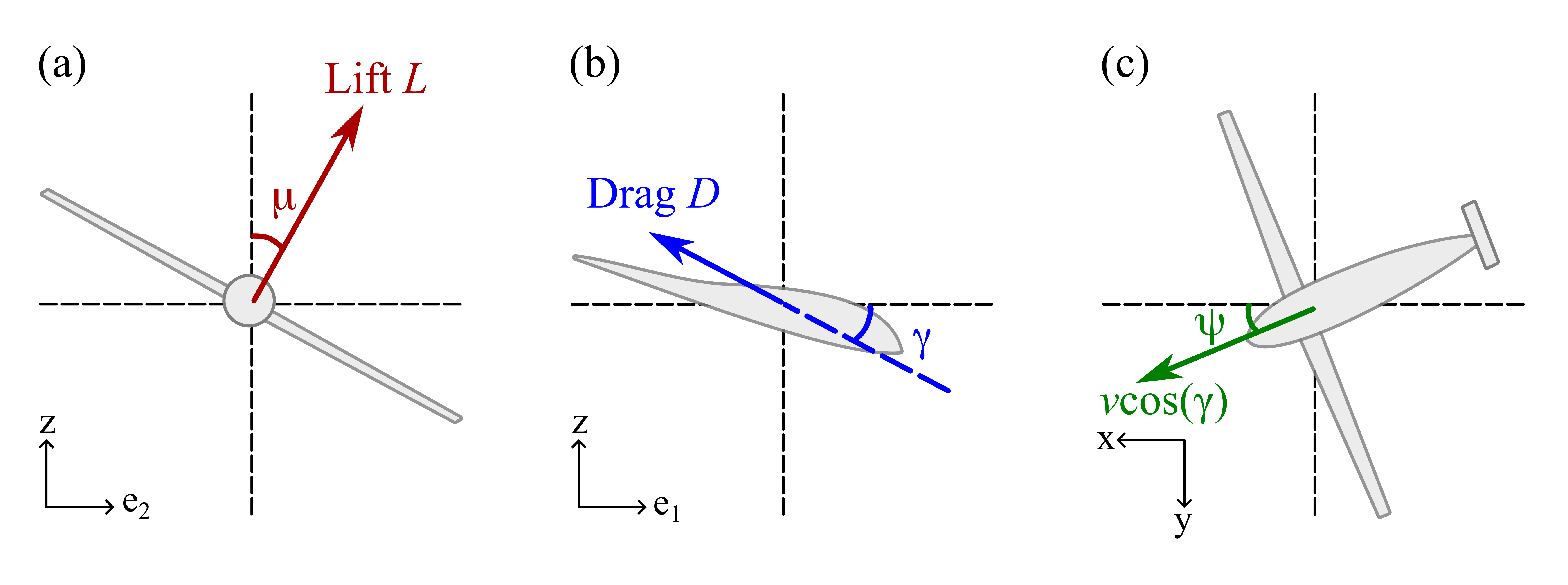}
  \end{center}
  \caption{Diagrams of the glider from several different directions using the global $(x,y,z)$ coordinate system and $(e_1,e_2)$ coordinate system defined in Fig.~\ref{fig:glider}. The angles determine the glider's orientation. (a) Bank angle $\mu$ is how much the glider rolls from the vertical. (b) Glide angle $\gamma$ is the angle between the heading and the horizontal. (c) Azimuth angle $\psi$ is the angle between the projection of the air velocity vector onto the horizontal plane and the $x$-axis.\label{fig:glider_angles} }
\end{figure}

As shown in Fig.~\ref{fig:glider_angles}, the glider's orientation can be
described using three angles. The bank angle $\mu$ sets the tilt of the wing
surfaces from vertical, and can be controlled by the glider itself. The glide
angle $\gamma$ measures the angle of the glider's motion from horizontal and is
given by $\gamma= \sin^{-1} v_z/v$. The azimuth angle measures the direction of
the glider in the horizontal plane and is determined so that $(v_x,v_y) =
(v\cos \gamma \cos \psi,v \cos \gamma \sin \psi)$.  With these definitions, the
glider's governing equations are given by
\begin{align}
  m \Dot{u}_x &= L \cos(\mu)\sin(\gamma)\cos(\psi)-L \sin(\mu)\sin(\psi)-D \cos(\gamma)\cos(\psi), \label{eq:gov1}\\
  m \Dot{u}_y &= L \cos(\mu)\sin(\gamma)\sin(\psi)+L \sin(\mu)\cos(\psi)-D \cos(\gamma)\sin(\psi), \\
  m \Dot{u}_z &= L \cos(\mu)\cos(\gamma)+D \sin(\gamma)-mg, \\
  \Dot{\br} &= \bu, \label{eq:gov4}
\end{align}
where $m$ is the mass of the glider and $g$ is the gravitational acceleration.
The energy is given by
\begin{equation}
  E= \frac{1}{2}mv^2  - m \vec{g} \cdot \vec{z}.
  \label{eq:energy}
\end{equation}
Taking the time derivative of Eq.~\eqref{eq:energy} and substituting in
Eqs.~\eqref{eq:gov1}--\eqref{eq:gov4} shows that the change in energy
is given by
\begin{align}
\dot{E} &=m\bv \cdot \dot{\bv} - m \bg \cdot \bu \notag \\
        &=m \bv \cdot (\dot{\bu}-\dot{\bw}) - m \bg \cdot (\bv+\bw) \notag \\
        &=\bv \cdot (m \dot{\bu} - m\bg) - m\bg \cdot \bw - m\bv \cdot \dot{\bw} \notag \\
        &=-vD+mgw_z-m\bv \cdot \dot{\bw},
\end{align}
which is equivalent to Eq.~\eqref{eq:energyrate} in the introduction. As
described in the introduction, the first term in this equation corresponds to
energy loss by drag. The other two terms in this equation correspond to energy
gain from altitude increase and energy gain from local wind gusts, both of
which are exploited by soaring birds \citep{taylor2016soaring}.

\subsubsection{Non-dimensionalized equations}
\label{sub:nondim}
Based on the glider's airspeed in steady flight, we define the glider's characteristic speed scale $v_c = \sqrt{2mg/\rho S}$, control time scale $ t_c = v_c/g$, and length scale $l_c=v_c^2/g$, where $\rho$ is the density of air and $S$ is the surface area of the wings. For soaring avian migrants, the typical gliding speed is $\SI{14}{m/s}$~\citep{tucker1970aerodynamics,horvitz2014gliding}; accordingly, in what follows we set $v_c = \SI{10}{m/s}$ and the typical control timescale $t_c = \SI{1}{s}$. We non-dimensionalize all velocities, lengths and times with respect to these three scales. Retaining the same notation, Eqs.~\eqref{eq:gov1}--\eqref{eq:gov4} become
\begin{align}
  \Dot{u}_x &= c_L v^2 \cos(\mu)\sin(\gamma)\cos(\psi)- c_L v^2 \sin(\mu)\sin(\psi)-c_D v^2 \cos(\gamma)\cos(\psi), \label{eq:ndgov1} \\
  \Dot{u}_y &= c_L v^2 \cos(\mu)\sin(\gamma)\sin(\psi)+ c_L v^2 \sin(\mu)\cos(\psi)-c_D v^2 \cos(\gamma)\sin(\psi), \\
  \Dot{u}_z &= c_L v^2 \cos(\mu)\cos(\gamma)+ c_D v^2 \sin(\gamma)-1,\label{eq:ndgov3} \\
  \Dot{\br}&=\bu. \label{eq:ndgov4}
\end{align}
We use Eqs.~\eqref{eq:ndgov1}--\eqref{eq:ndgov4} in the numerical
implementation. The non-dimensionalized energy is given by dividing
Eq.~\eqref{eq:energy} by a factor of $m v_c^2$, so that
\begin{equation}
  \epsilon =\frac{1}{2}v^2+z.
  \label{eq:ndenergy}
\end{equation}
The non-dimensionalized rate of energy gain is
\begin{equation}
\dot{\epsilon}=-c_D v^3+w_z-\bv \cdot \dot{\bw},
\label{eq:ndenergy_gain}
\end{equation}
with matches Eq.~\eqref{eq:energyrate}.

\subsection{Simulations of turbulent flow}\label{sec:wind}
We introduce a kinematic model for the wind field that the glider experiences,
which captures the broad range of length scales and timescales characteristic
of turbulent flows. For the three-dimensional wind field $\bw(\br,t)$, at
position $\br$ and time $t$, we consider a homogeneous and isotropic turbulent
flow that can be described by a Fourier transform,
\begin{equation}
\bw(\br,t)=(2\pi)^{3/2}\int_{-\infty}^{\infty} d^3 \bk e^{i\bk \cdot \br} \tilde{\bw}(\bk,t),
\end{equation}
where $\Tilde{\bw}(\bk,t)$ is the complex-valued amplitude of the wave with
wave number $\bk$ at time $t$. We simulate a turbulent field that reproduces
the second-order Kolmogorov statistics \citep{fung1992kinematic} where the energy contained in waves of magnitude $k=\|\bk\|$ scales like $E(k) \sim k^{-5/3}$. In our numerical implementation, using the non-dimensionalization introduced in Sec.~\ref{sub:nondim}, the wind velocity is computed in a periodic box $[0,B)^3$ using a discrete $N\times N \times N$ complex Fourier modes $\tbw_{\alpha\beta\zeta}$. Define a lattice spacing $h=B/N$. For a lattice point $\br=(hj,hl,hq)$, the wind velocity is given by the discrete Fourier transform
\begin{equation}
  \bw(\br,t) = \sum_{\alpha=0}^{N-1} \sum_{\beta=0}^{N-1} \sum_{\zeta=0}^{N-1} \tbw_{\alpha\beta\zeta}(t) e^{2\pi i (j\alpha + l\beta + q\zeta)/N}.
  \label{eq:dft}
\end{equation}
We treat the complex Fourier modes as indexed periodically, so that
$\tbw_{(\alpha+N),\beta,\zeta} = \tbw_{\alpha\beta\zeta}$, with similar
relations for $\beta$ and $\zeta$.  In order to ensure that $\tbw(\br,t)$ is
real-valued, the discrete Fourier coefficients must satisfy
\begin{equation}
  \tbw_{\alpha\beta\zeta}=\bar{\tbw}_{-\alpha,-\beta,-\zeta}.
  \label{eq:mode_constr}
\end{equation}
Due to the periodicity of the complex Fourier modes, the sums in
Eq.~\eqref{eq:dft} can be reordered. Define $M=\lfloor \tfrac{N}2 \rfloor$, and
\begin{equation}
  f(n) = \begin{cases}
    \tfrac12 & \qquad \text{if $|n| = N/2$,} \\
    1 & \qquad \text{if $|n| < N/2$.}
  \end{cases}
\end{equation}
Then for a lattice point $\br=(hj,hl,hq)$, Eq.~\eqref{eq:dft} is equivalent to
\begin{equation}
  \bw(\br,t) = \sum_{\alpha=-M}^M \sum_{\beta=-M}^M \sum_{\gamma=-M}^M f(\alpha) f(\beta) f(\zeta) \tbw_{\alpha\beta\zeta}(t) e^{2\pi i (j\alpha + l\beta + q\zeta)/N}.
  \label{eq:dft2}
\end{equation}
When $N$ is odd, each sum has exactly $N$ terms. When $N$ is even, each sum has exactly $N+1$ terms, but the $f$ function causes the two extremal indices at $\pm M$ to be counted with half weighting. Those two extremal indices have the same Fourier mode coefficient, due to the periodicity. Thus in both cases the sums evaluate the same terms as in Eq.~\eqref{eq:dft}.

Equations \eqref{eq:dft} \& \eqref{eq:dft2} are equivalent at lattice points, but this relies on cancellation of factors of $e^{2\pi i}$. At off-lattice points they are not the same, since Eq.~\eqref{eq:dft} involves a sum over rapidly oscillating exponentials. By contrast, Eq.~\eqref{eq:dft2} provides a representation of the Fourier exponentials in the lowest frequencies. Since the gliders move across arbitrary locations, we therefore make use of Eq.~\eqref{eq:dft2} to evaluate the wind field at any position $\br$.

\begin{figure}
  \centerline{\includegraphics[width=10cm]{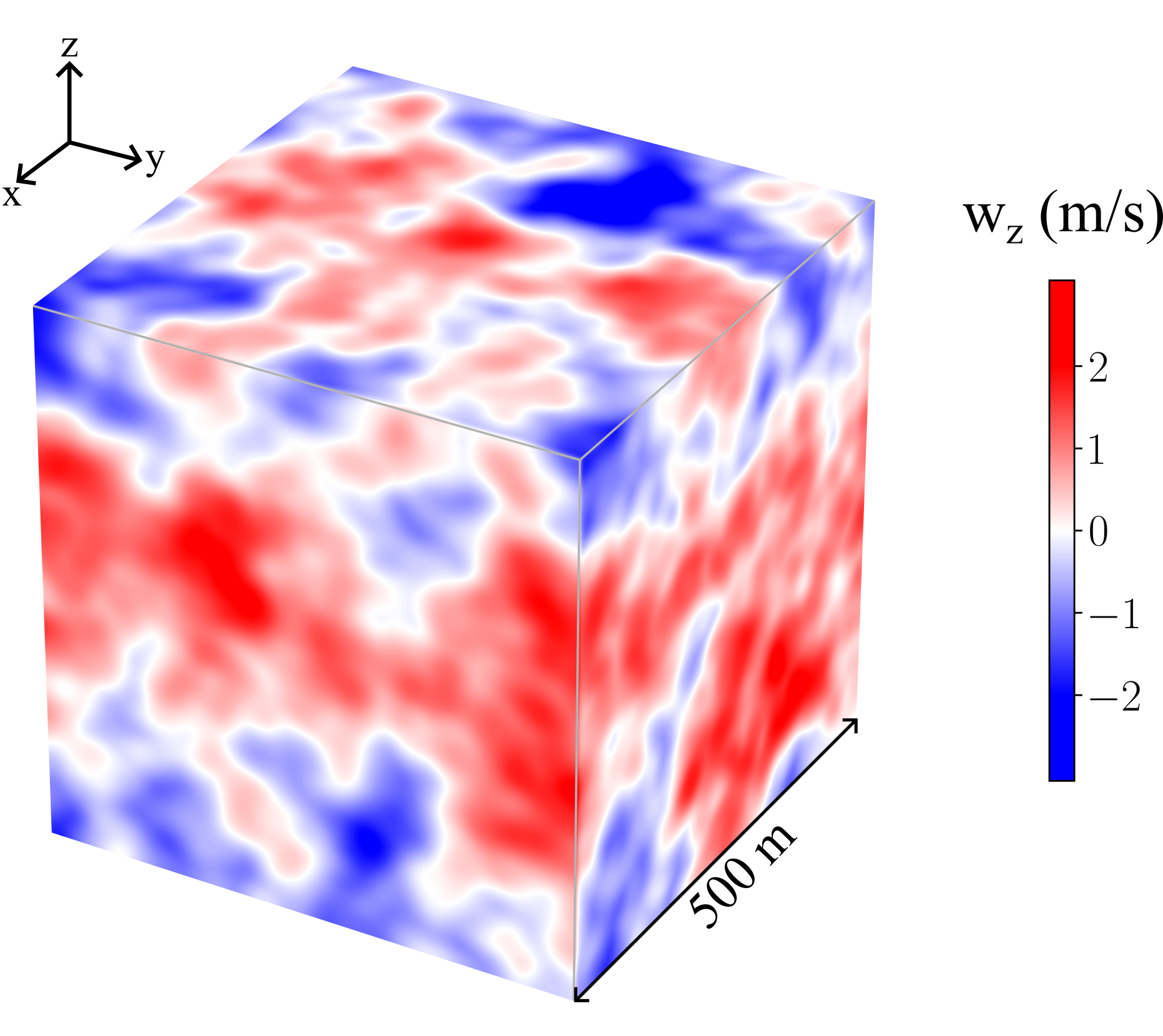}}
  \caption{The numerical simulation of the three-dimensional homogeneous and isotropic turbulence. The figure shows vertical wind velocity field. Red and blue indicates large ascending and descending currents respectively. }
  \label{fig:wind}
\end{figure}

We now identify $\tbw(\bk) = \tbw_{\alpha\beta\zeta}$ for $\bk=(2\pi
\alpha/B,2\pi \beta/B,2\pi \zeta/B)$. Following the model of
\citet{fung1992kinematic} we assume that each Fourier mode evolves as an
Ornstein--Uhlenbeck process with time scale $\tau(k)$ so that
\begin{equation}
 d \tbw(\bk,t)=-\tbw (\bk,t)\frac{dt}{\tau(k)} + a(k) dW(t),
 \label{eq:sde}
\end{equation}
where $W(t)$ denotes the Wiener process. To match Kolmogorov scaling, we
require $\tau(k) \sim k^{-2/3}$. $a(k)$ is set such that the energy spectrum
$E(k) \sim k^{-5/3}$. The specific expressions for $a(k)$ and $\tau(k)$ are
given in Sec.~\ref{sec:covariance}. Figure \ref{fig:wind} shows an example of
the simulated wind field, where we use a cubic domain with side length
$B=\SI{500}{m}$ for the fluid field, with periodic boundary conditions. The fluid
is simulated using $64^3$ modes. The root-mean-square (RMS) wind speed is set
to be $\SI{0.5}{m/s}$ based on typical amplitudes of atmospheric wind
fluctuations. Appendix \ref{app:wind_num_imp} describes how the wind field is
computed efficiently using multi-threaded programming.

\subsection{Derivation of the spatio-temporal covariance}
\label{sec:covariance}
In the later sections, we are going to infer the wind information at unknown locations and times with the partial knowledge of the field. Thus, we need to know the correlation of the wind function. The spatial domain is $\Omega = [0,B)^3$. To compute the spatio-temporal covariance function, we first write
 \begin{align}
\langle \bw(0,0), \bw(\br,t)\rangle &=  (2\pi)^{-3} \int_{\tilde{\Omega}} d^3 \mathbf{k}\, \int_{\tilde{\Omega}} d^3 \mathbf{k'}\, e^{i\mathbf{k' \cdot r}}\langle \tbw(k,0),\tbw(k',t)\rangle \nonumber \\
 &=  (2\pi)^{-3} \int_{\tilde{\Omega}} d^3 \mathbf{k}\, \int_{\tilde{\Omega}} d^3 \mathbf{k'}\, e^{i\mathbf{k' \cdot r}} (12
 \delta(\bk+\bk')e(k)e^{-|t|/\tau(k)}
\end{align}
 where $\tilde{\Omega}$ is the reciprocal space and the second step comes from
 \begin{align}
 \langle \tbw(\bk,0),\tbw(\bk',t)\rangle &= \langle \tbw_R(\bk,0),\tbw_R(\bk',t)\rangle - \langle \tbw_I(\bk,0),\tbw_I(\bk',t)\rangle \nonumber \\
 &+ i\langle \tbw_R(\bk,0),\tbw_I(\bk',t)\rangle +i \langle \tbw_I(\bk,0),\tbw_R(\bk',t)\rangle \nonumber \\
 &=(\delta(\bk-\bk')+\delta(\bk+\bk'))6e(k)e^{-|t|/\tau(k)} \nonumber \\
  &-(\delta(\bk-\bk')-\delta(\bk+\bk'))6e(k)e^{-|t|/\tau(k)}+0+0 \nonumber \\
  &=12 \delta(\bk+\bk')e(k)e^{-|t|/\tau(k)}.
\end{align}
Integrating out $\bk'$ gives
\begin{align}
\langle \bw(0,0), \bw(r,t)\rangle &=  \frac{3}{2\pi^3} \int d^3 \mathbf{k} e(k) e^{-i\mathbf{k \cdot \br}} e^{-|t|/\tau(k)}  \nonumber \\
 &\approx  \frac{3}{\pi^2} \int_{-1}^1 d(\cos\theta)\int_{\kmin}^{\kmax}dk \, k^2 e(k)e^{ikr \cos\theta}e^{-|t|/\tau(k)},
\end{align}
where we have approximated the integral over the box by an integral over a spherical shell as $k^2e(k)$ decays to zero for large $k$. The characteristic timescale $\tau(k) =  \tau_c (l_c k)^{-2/3}$ has prefactor $\tau_c$ that sets the overall timescale of temporal fluctuations. The smallest wave number is $\kmin=\sqrt{3}\pi/B$. The largest wave number is determined by the smallest resolved length scale, which is $1/N$ in an $N^3$ grid, therefore, we have the largest wave number $\kmax=\sqrt{3}\pi/(B/N)=\sqrt{3}\pi N/B$. Let $k^2e(k)=\alpha k^{-5/3}$ where $\alpha$ will be set such that the variance of each component is 1. Then we have
\begin{align}
\langle \bw(0,0), \bw(0,0)\rangle =1 &\approx \frac{3}{\pi^2} \int_{-1}^1 d(\cos\theta)\int_{\kmin}^{\kmax}dk \,\alpha k^{-5/3} \nonumber\\
&= \frac{6\alpha}{\pi^2} \int_{\kmin}^{\kmax}dk \,k^{-5/3} \nonumber \\
&=\frac{9\alpha}{\pi^2}(\kmin^{-2/3}-\kmax^{-2/3}),
\end{align}
which gives
\begin{align}
    \alpha&\approx \frac{\pi^2}{9}(\kmin^{-2/3}-\kmax^{-2/3})^{-1}, \nonumber \\
    e(k)=\alpha k^{-11/3}&=\frac{\pi^2}{9}k^{-11/3}(\kmin^{-2/3}-\kmax^{-2/3})^{-1}, \nonumber \\
    a(k)=\sqrt{4e(k)/\tau(k)}&=\frac{2}{3}3^{-1/6}\pi^{1/3}k^{-3/2}\sqrt{(\kmin^{-2/3}-\kmax^{-2/3})^{-1}}.
\end{align}
The covariance function for each component is $K(r,t)=\langle \bw(0,0), \bw(\br,t)\rangle$, and
\begin{equation}\label{eq:covariance_fun}
    K(r,t)=\frac{6\alpha}{\pi^2}\int_{\kmin}^{\kmax}dk \,k^{-5/3} \sinc(kr)e^{-|t|/\tau(k)}.
\end{equation}
Due to the translation invariance of the wind field, for any two locations $\br_1, \br_2$, times $t_1, t_2$, we compute $r=\|\br_1-\br_2\|, t=|t_1-t_2|$ and apply Eq.~\eqref{eq:covariance_fun} to get their correlation.

\begin{figure}
    \begin{center}
    \includegraphics[width=\textwidth]{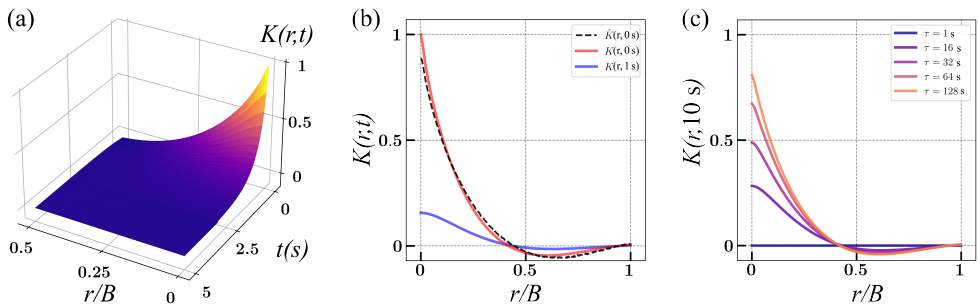}
    \end{center}
    \caption{Illustration of the kernel function. (a) The correlation function with respect to time $t$ and distance $r$ when $\tau_c=\SI{1}{s}$. (b) The correlation function with respect to $r$ at $t=0$ and $t=1$ when $\tau_c=\SI{1}{s}$. The correlation of the simulated wind field $\hat{K}(r,t)$, which is computed using 128 wind fields, each has 1024 samples at random locations. (c) The correlation function with respect to $r$ at $t=\SI{10}{s}$ when $\tau_c=[\SI{1}{s},\SI{16}{s},\SI{32}{s},\SI{64}{s},\SI{128}{s}]$.}
    \label{fig:kf}
\end{figure}

The statistical properties of the flow are determined by the stationary, isotropic covariance kernel of each wind velocity component as shown in Eq.~\eqref{eq:covariance_fun}. The analytical correlation function $K(r,t)$ is illustrated in Fig.~\ref{fig:kf}, where the correlation decreases exponentially as difference in time or distance increases. There is significant correlation when $r/B<0.5$ and $t<1$. The correlation of the wind field simulation $\hat{K}(r,t)$ matches the analytical values $K(r,t)$.

\section{Implementation of the glider's navigational strategy}
\subsection{Overview of the navigational strategy}
We now describe the navigational strategy, which maps the glider's wind velocity and positional history to one of three bank angle changes, namely, $\Delta \mu = 0^{\circ},\pm 10^{\circ}$ to a maximum (minimum) of $\mu = \pm 40^{\circ}$. A glider with memory size $M$ and history $H_t$ at time $t$ contains the glider's past locations and wind measurements collected at fixed intervals $\Delta t = t_c/2$, that is, \smash{$H_{t} = \bm{r}_{t-M\Delta t} \bm{w}_{t-M\Delta t}\bm{r}_{t- (M-1)\Delta t} \bm{w}_{t- (M-1)\Delta t} \hdots \bm{r}_t \bm{w}_t$}. The glider chooses an action so as to maximize the total expected energy gained in time $d\Delta t$, $\langle \varepsilon_{t + d\Delta t} \rangle_{H_t} - \varepsilon_t$, where $d\Delta t  = 10 t_c$ sets the planning horizon. This maximal expected energy gained given a particular history, $V(H_t)$, satisfies the recursive Hamilton--Jacobi--Bellman equation~\citep{bardi1997optimal},
\begin{equation}\label{eq:bellman}
    V(H_t) = \max_{a \in \mathcal{A}} \hspace{0.125cm}  \langle \varepsilon_{t+\Delta t}  - \varepsilon_t +  V(H_{t+\Delta t}) \rangle_{H_t,a},
\end{equation}
where the expectation is over wind configurations encountered along the glider's trajectory from $t$ to $t+\Delta t$ with boundary condition $V(H_{t+d \Delta t}) = 0$. The optimal action is the one that attains the maximum over the set $\mathcal{A}$ of all possible actions that the glider can take. The expectation implicitly contains the `propagator', i.e., the conditional probability density of the wind velocity at a new location, $P(\bm{w}_{t'} | H_t,\bm{r}_{t'})$ for $t' > t$. Computing the optimal action is generally challenging due to the evaluation of the exponential number ($\sim |\mathcal{A}|^d$) of possible future paths and the wind configurations encountered along these paths. We now describe three simplifications made to derive an efficient, online algorithm from the general expression Eq.~\eqref{eq:bellman}.

First, we observe that the glider's future trajectory and energy gain have approximately a linear dependence on $\bm{w}$ if $w$ is much smaller than the typical airspeed of the glider ($w \ll v_c$). Since the flow is defined by its second-order statistics in Eq.~\eqref{eq:covariance_fun}, it is a Gaussian process. We perform Gaussian process regression to predict wind velocities in future time and locations with given history $H_t$, and then we compute the expected energy gain along a large number of paths (here, $10^4$ paths).

Second, we numerically evaluate the optimal path using Monte Carlo sampling. However, the large number of future paths remains a challenge and efficient pruning techniques are required to make planning tractable. To prune sub-optimal paths during planning, we implement a Monte Carlo tree search (MCTS)~\citep{browne2012survey}. MCTS uses a tree search algorithm to balance exploration and exploitation of future paths. The algorithm stochastically samples different paths, and at each branch chooses an action based on the expected energy gain. The sampling is biased toward paths with higher energy gain, but with an additional exploration bonus to ensure less promising paths still have a chance of being explored.

Finally, we employ MCTS to find the sequence of actions $a_1,a_2,\dots,a_d$ that approximately maximize the expected energy gain, i.e., $V(H_t) \approx \max_{a_1,a_2,\dots,a_d} \langle \varepsilon_{t + d\Delta t} - \varepsilon_t \rangle_{H_t,a_1,a_2,\dots,a_d}$. Note that this is not equivalent to optimizing Eq.~\eqref{eq:bellman} as the max operator is over paths whose expected energy gain is computed given $H_t$.

In the following subsection, we provide full details of the Gaussian process regression (GPR) that the glider can use to predict the wind field based on its history of observations. Then we describe the Monte Carlo tree search (MCTS), which the glider uses to control its flight to maximize energy gain. Finally, we summarize the setup of the simulation process.

\subsection{Gaussian process regression}
\label{sec:gpr}
\begin{figure}
    \begin{center}
    \includegraphics[width=\textwidth]{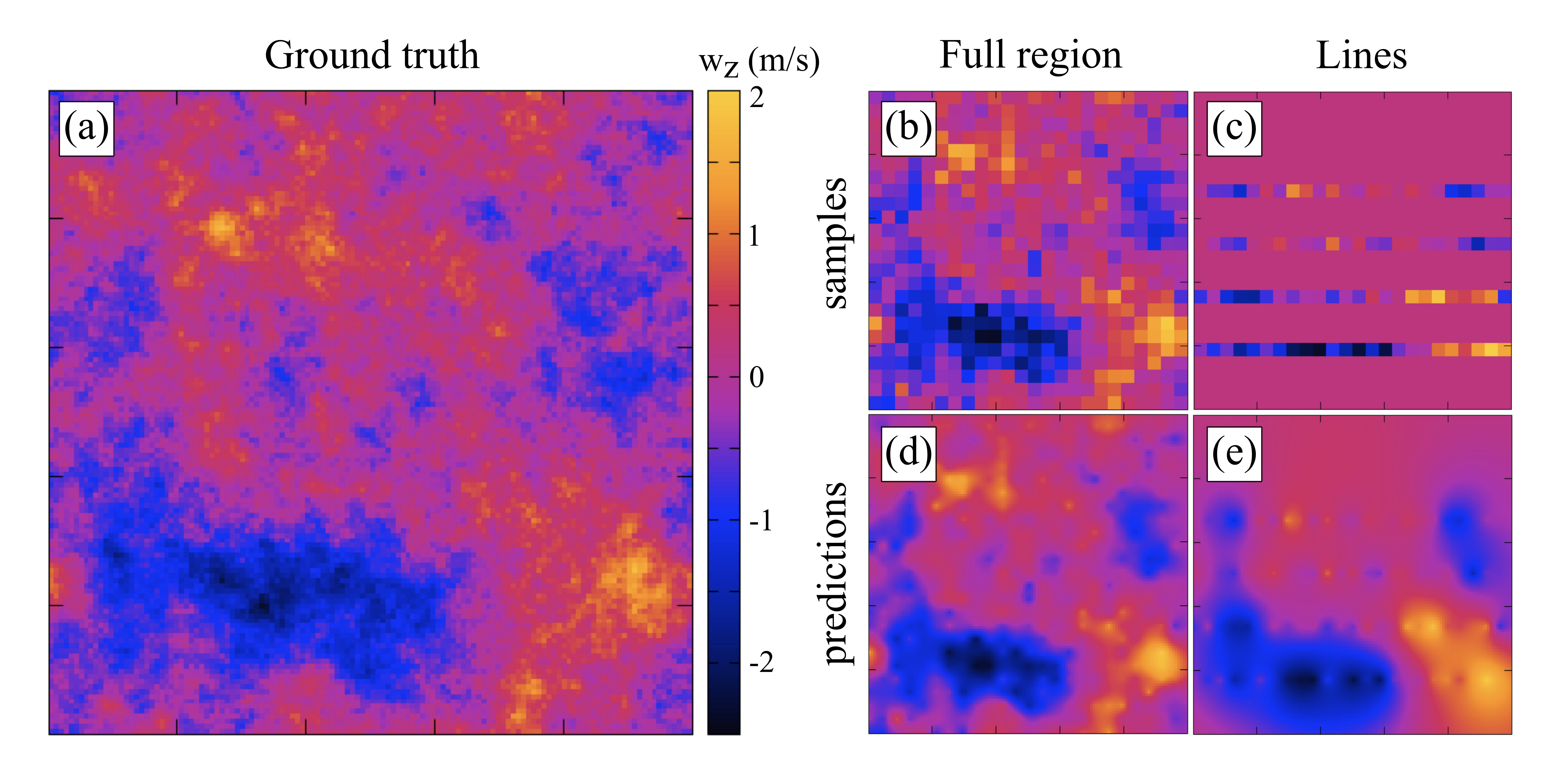}
    \end{center}
    \caption{Illustration of GPR prediction. (a) Vertical wind velocity in $128\times128$ $xz$-plane computed using the FFTW library as described in Sec.~\ref{sec:wind} and Appendix~\ref{app:wind_num_imp}. (b) Coarse $24\times 24$ samples of the wind field (c) Four lines of samples of the wind field on a $4\times24$ grid. (d) The corresponding GPR predictions of $w_z$ using the samples in (b). (e) The corresponding GPR predictions using the samples in (c).\label{fig:gpr_sample}}
\end{figure}

\begin{figure}
    \begin{center}
       \includegraphics[width=\textwidth]{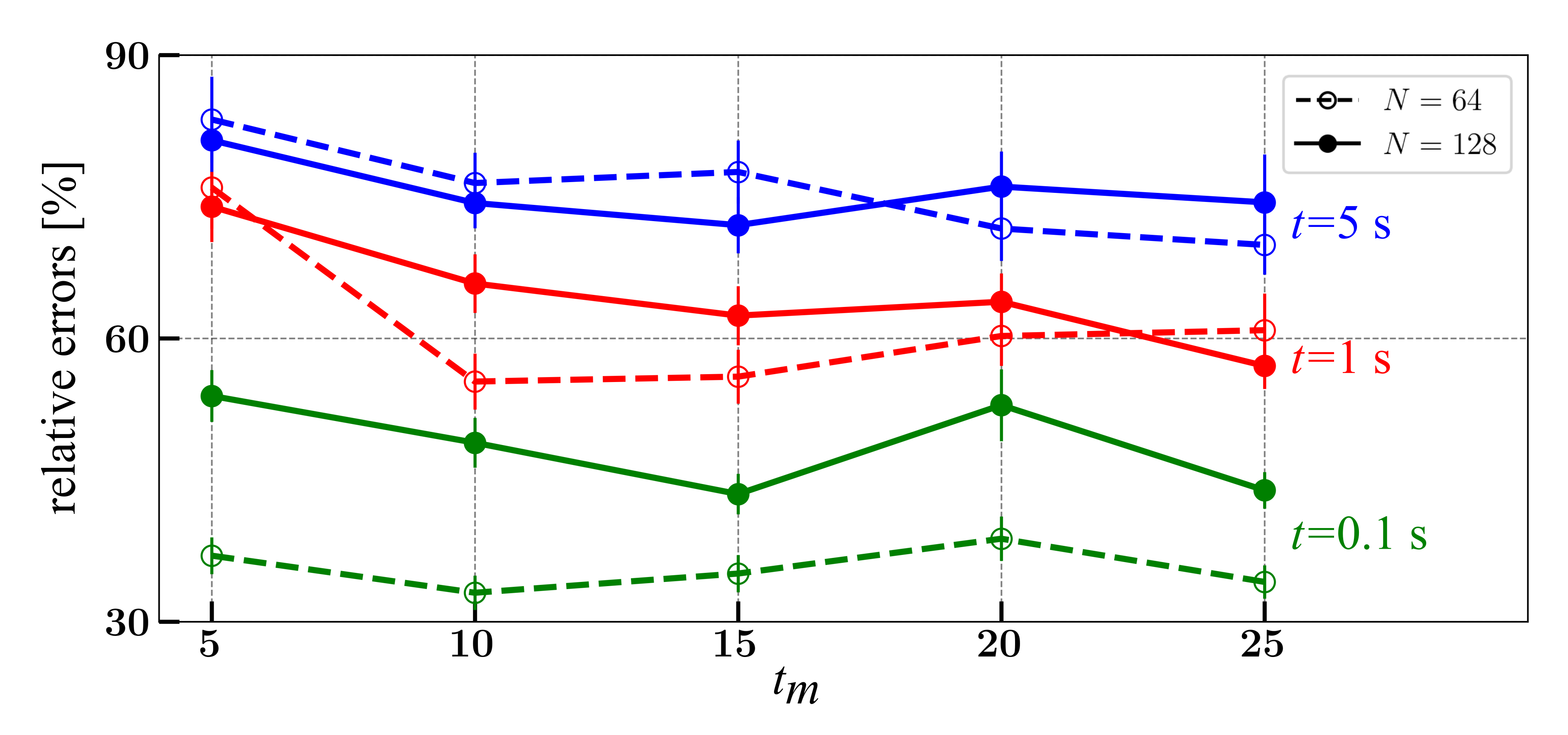}
    \end{center}
    \vspace{-0.3em}
    \caption{The relative errors of wind prediction on the frozen wind field in the future time. There are 200 gliders randomly travel in the frozen wind field. In the first \SI{25}{s}, they collect information and store in the memory with $M=50$. Then they use GPR to do predictions at their current locations after \SI{0.1}{s}, \SI{1}{s} and \SI{5}{s}. The figure shows the relative errors between the wind prediction $\mathbf{\hat{w}}$ and actual wind velocities $\mathbf{w}$.}
    \label{fig:gpr_pred}
\end{figure}

We define the mean function $m(\bx)$ and the covariance function $K(\bx,\bx')$ of a Gaussian process $f(\bx)$ as~\citep{wan2017reduced}
\begin{align}
    m(\bx)&=E[f(\bx)], \\
    f(\bx) & \sim GP( m(\bx),K(\bx,\bx')).
\end{align}
Suppose we have an observed dataset $X$ of size $M$, and define $f=f(X)$. Let $f^*$ be unobserved data at $X^*$. Then the joint distribution of $\bm f$ and $\bm f^*$ is
\begin{equation}
  \begin{bmatrix}
\bm f\\
\bm f^*
\end{bmatrix} \sim N \left(
\begin{bmatrix}
m(X)\\
m(X_*)
\end{bmatrix},\begin{bmatrix}
K(X,X) & K(X,X^*)\\
K(X^*,X) & K(X^*,X^*)
\end{bmatrix}
\right)
\end{equation}
Conditioning the joint Gaussian prior distribution on the observations, we have
\begin{equation}
        \bm f^*|\bm f,X,X^* \sim N(K(X^*,X)K(X,X)^{-1}\bm f, K(X^*,X^*)-K(X^*,X)K(X,X)^{-1}K(X,X^*)).
        \label{eq:posterior}
\end{equation}
Since our wind components are generated from many random modes that evolve as a Gaussian process, we use Gaussian process regression to perform wind inference, that is, we can apply Eq.~\eqref{eq:posterior} to perform wind predictions using a history of observations. We calculate the kernel function directly from the wind model Eq.~\eqref{eq:covariance_fun}. Figure \ref{fig:gpr_sample} gives an example of the wind predictions with sparse sampling in the wind field. It recovers the large scale features in the wind field. To validate the accuracy of the GPR prediction, we set 10 frozen wind fields, and the let 20 gliders randomly explore for $t_m=$\SI{25}{s}, with the glider storing information $(\bw,\br,t)$ in memory every $\Delta t = \SI{0.5}{s}$. When the memory is full, we perform predictions after 0.1, 1, and \SI{10}{s}, where \SI{10}{s} is the planning horizon of gliders. Figure \ref{fig:gpr_pred} demonstrates that the GPR prediction is accurate over short distances and times, and has large errors over long times where the correlation is weak. Increasing the duration of memory collection helps improve the performance of the GPR prediction. Increasing the number of modes brings more fine details into the wind field, which makes the prediction harder.

Gaussian process regression has the potential to be computationally expensive, since evaluating the mean in Eq.~\eqref{eq:posterior} requires computing $K(X,X)^{-1}$, which changes on each step. We therefore accelerate the computation of these inverses, as described in Appendix \ref{app:epgr}.

\subsection{Monte Carlo tree search}\label{sec:mcts}
Monte Carlo tree search is an efficient method for planning into the future and making decisions. The key idea is to randomly build a search tree of possible actions into the future. In the method, many random play-outs of the search tree are considered, where more promising branches are visited more often, but there is still exploration of less fruitful options. After each play-out, the final reward is used to update the rewards at each node along the path of that play-out. After playing out many simulations, the tree provides an estimation of the best trajectory.Energy-positive soaring using transient turbulent fluctuations

Define the sequence of $k$ actions $A_k=a_1 a_2 a_3 \cdots a_k$, where each action is a change in bank angle. Specifically each $a_j$ will be $\Delta \mu=\pm 10^\circ$, and $\mu$ ranges from $-40^\circ$ to $40^\circ$. The value function $V(A_k)$ denotes the estimated average energy gain from $A_k$ till terminate. $N(A_k,a_{k+1})$ denotes the number of visited times of $a_{k+1}$ from $A_k$. For node $i$ of the tree, it stores the information of the action sequence $A_i$, the value function $V(A_i)$ and the total number of visits $N(A_{i-1},a_i)$. The UCT algorithm selects the action that maximizes the upper confidence bound \citep{kocsis2006bandit}
\begin{equation}
     U(A_k,a_{k+1})=V(A_k\cup a_{k+1})+2C\sqrt{\frac{2\ln(\sum_a N(A_k,a))}{N(A_k,a_{k+1})}}.
\end{equation}
The first term in the UCT bound encourages choosing high-value action, while the second term encourages exploring the less visited actions. $C=0.5$ is the exploration parameter that can be varied if more exploration is desired.

In one Monte Carlo simulation, a sequence of actions is selected using the UCT algorithm, then the energy gain is estimated using Eq.~\eqref{eq:ndenergy_gain} and propagated back to update $V(A_k)$. To estimate the energy gain, the wind velocities in future times and locations are needed. We perform the GPR prediction to estimate wind and then compute the estimated energy gain. A sufficient amount of Monte Carlo simulations are played out and the action that has the highest value is chosen to be the actual action.

\begin{figure}
    \centering
    \includegraphics[width=10cm]{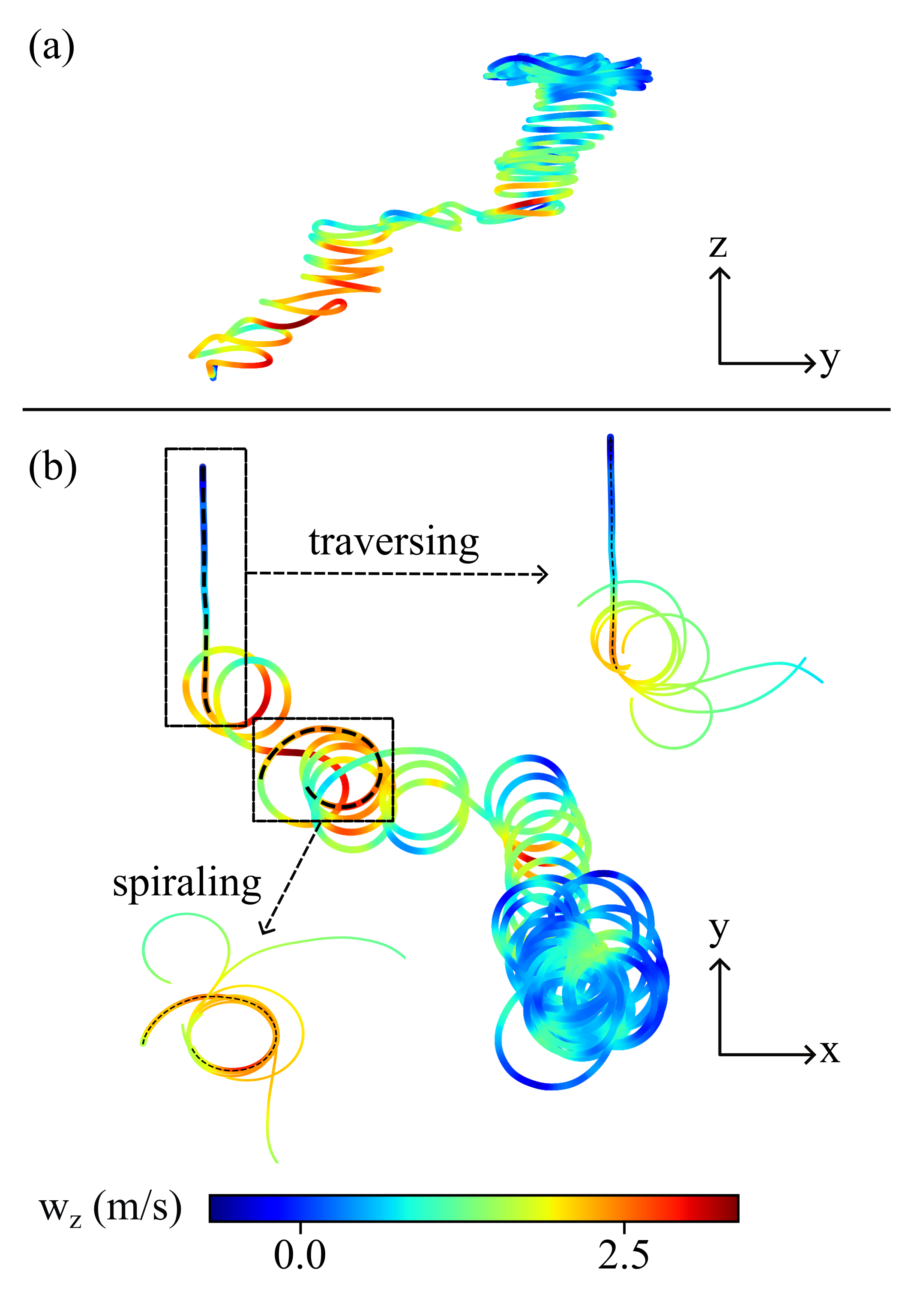}
    \vspace{-0.5em}
    \caption{The glider's trajectory in a frozen wind field. The color shows the vertical wind velocity that applied on the glider. (a) The glider's trajectory in the $yz$-plane. (b) The glider's trajectory in the $xy$-plane. The glider exhibits `traversing' behavior at the beginning of the searching and a switch to `spiraling' behavior once an ascending current has been found. The zoomed-in figures show some examples of the MCTS paths.}
    \label{fig:gl_traj}
\end{figure}

\section{Results and Discussion}
We simulate gliders soaring in the wind field. They collect information during exploration, and then use MCTS planning scheme with GPR prediction to choose the optimal action and move forward. The gliders may know the full region wind field or only the part they have experienced. We define them as a \textit{full information glider} and \textit{partial information glider} accordingly.  

\subsection{Glider setup}
We define the full information glider to be the glider that knows the wind velocities in the whole region at the current time. In the MCTS planning part, it uses the precise wind velocities $\bw(\br,t)$ in the current time. When planning, it uses $\bw(\br,t)$ to predict $\bw(\br,t+\Delta t)$ with Eq.~\eqref{eq:kt}, which only involves temporal correlations in the wind field, so that
\begin{equation}\label{eq:kt}
  K(0,t)=\frac{|t|}{\tau(\kmin)-\tau(\kmax)}\left(\Gamma\left(-1,\frac{|t|}{\tau(\kmin)}\right)-\Gamma\left(-1,\frac{|t|}{\tau(\kmax)}\right)\right).
\end{equation}

The partial information glider refers to the glider that only knows the wind velocities at its visited locations. It stores limited wind information with memory size $M={10,20,30,40,50}$. In the planning part, it uses GPR to predict the wind at unknown locations and times using the prior knowledge in memory.

\subsection{Gliders in the static wind field}
We perform 10 simulations of the wind field, each containing 20 gliders, making 200 gliders in total. We simulate for a duration of \SI{500}{s}. We first consider gliders navigating in wind fields that remain constant in time, which have spatial correlations specified by $K(r,0)$ in Eq.~\eqref{eq:covariance_fun}. We can customize MCTS by choosing the depth of the planning tree and number of simulated paths. The planning horizon is chosen to let the glider sufficiently explore the field. We set the planning horizon to be \SI{10}{s} for two reasons: first, the planning horizon is sufficient for the glider to make a full circle to explore the field; second, there is reasonable correlation of wind within \SI{10}{s} (Fig.~\ref{fig:kf}). As shown in Fig.~\ref{fig:gpr_pred}, the gliders cannot gather accurate information in after \SI{5}{s}. We set the number of paths to be $10^4$, which is sufficient to select the optimal action and is computationally efficient. In Fig.~\ref{fig:gl_traj}, we show a sample path of the glider along with the trajectories explored at a single instance of MCTS along its path. The glider exhibits distinct bouts of localized spiraling behavior during which they gain height, similar to soaring patterns observed during thermal soaring. These bouts of spiraling punctuated by flat traversals can be intuitively viewed as `foraging' behavior: the ascending currents that drove spiraling behavior expire when the height gained in these bouts exceeds the typical correlation length ($\simeq \SI{50}{m}$). Subsequently, the glider traverses less valuable regions of the flow towards new energy-rich regions.

\begin{figure}
  \centering
    \includegraphics[width=\textwidth]{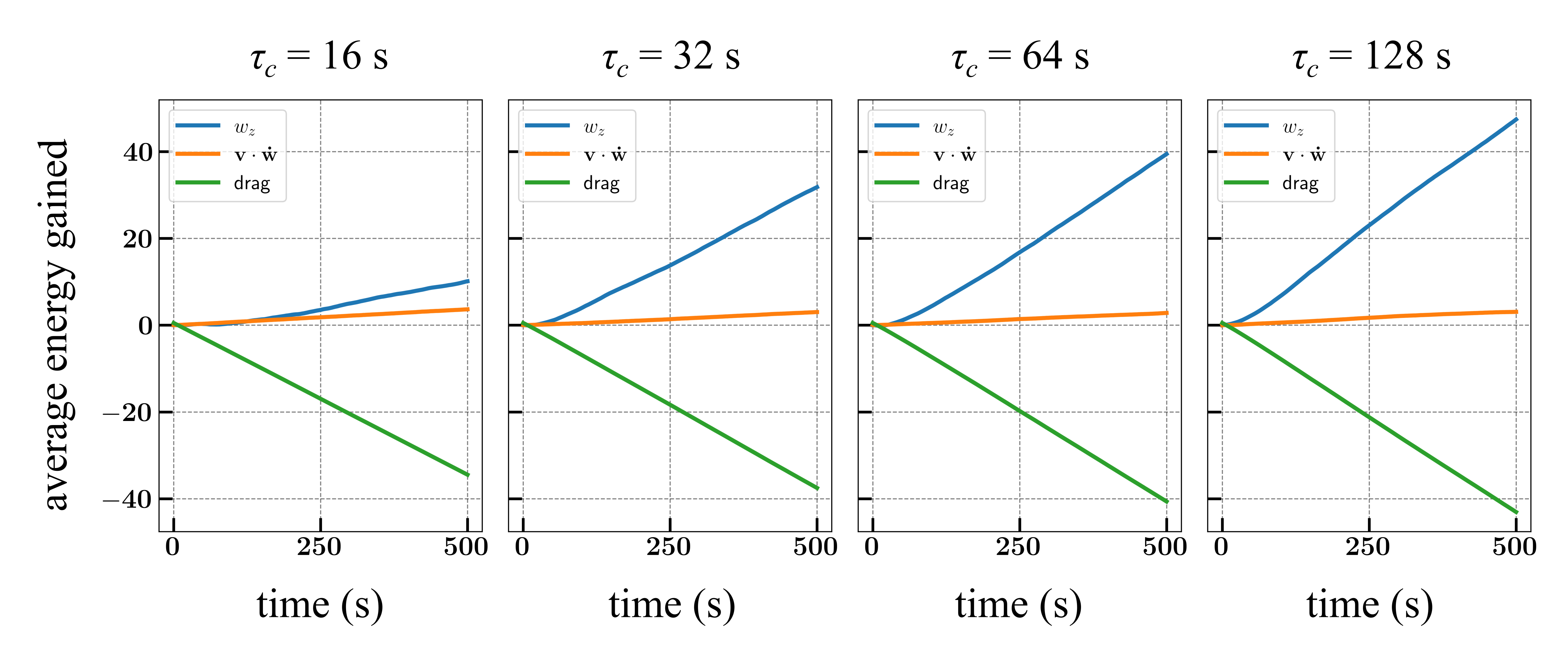}
    \vspace{-0.4em}
    \caption{The components of average energy gained. We simulated 200 gliders for $\tau_c=16, 32, 64, \SI{128}{s}$ with $M=40$ and show the average energy gained in Eq.~\eqref{eq:ndenergy_gain} separately. The $w_z$ term dominates the energy contribution. }
    \label{fig:energy_comp}
\end{figure}

Over time scales much longer than than $\SI{1}{s}$, since $v$ remains comparable to $v_c$, energy extracted from the flow is primarily converted to potential energy. To investigate the composition of energy gain, we compute each component in Eq.~\eqref{eq:ndenergy_gain}. Figure \ref{fig:energy_comp} shows that the contributions from wind fluctuations are insignificant. The energy gained is mainly due to upcurrents.

\begin{figure}
\centering
\includegraphics[width=7.4cm]{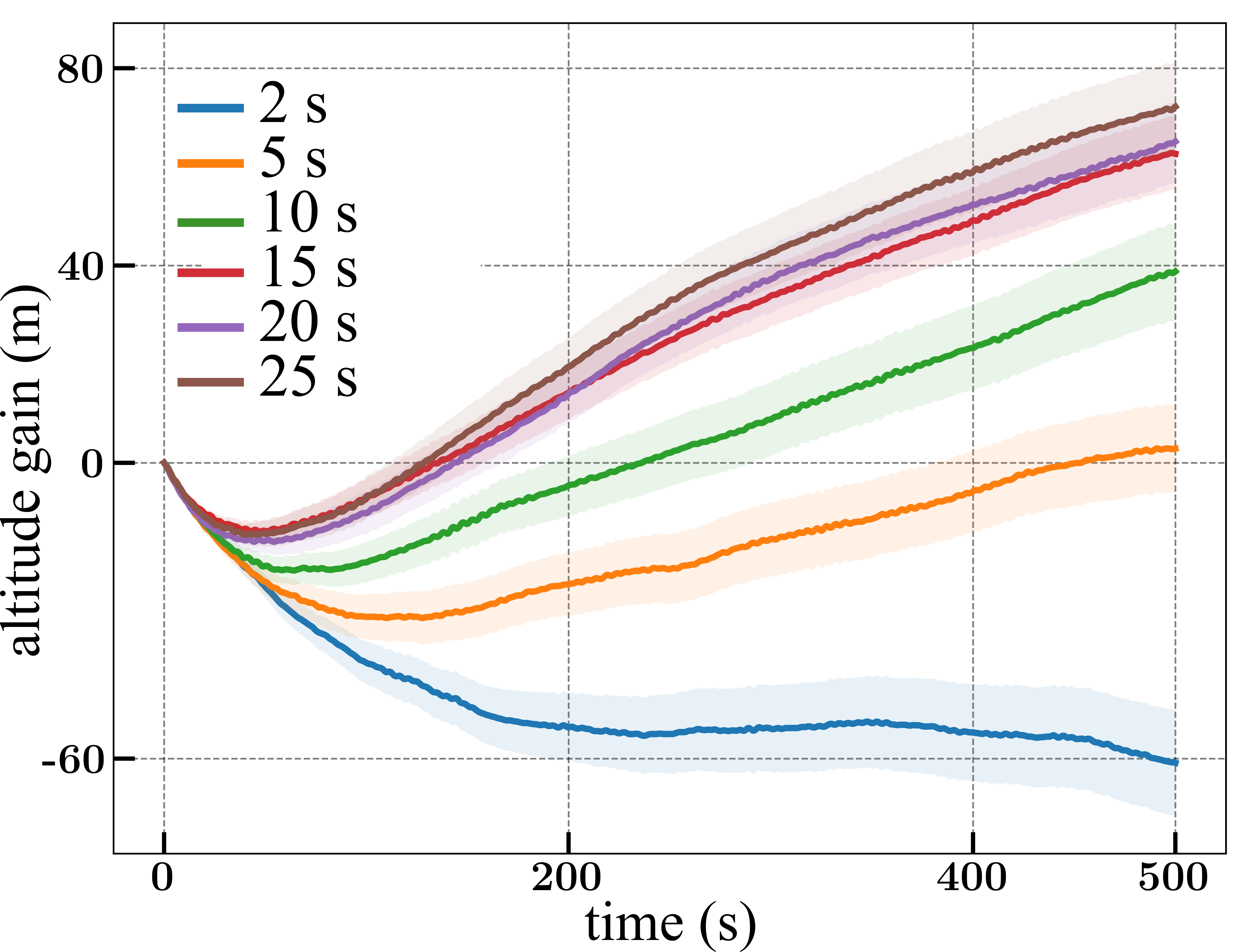}
\caption{Gliders' altitude gain in a frozen wind field, using the partial information acquired from the measured wind histories, using different memory durations $t_m=[\SI{2}{s},\SI{5}{s},\SI{10}{s},\SI{15}{s},\SI{20}{s},\SI{25}{s}]$. All the gliders initially descend, but start to gain altitude after building a better model for the wind field. The slope of the altitude gain from \SI{200}{s} onward is measured as the climb rate. The climb rate measurement is defined in Eq.~\eqref{eq:effic}.\label{fig:frozen}}
\end{figure}

We thus use the climb rate $c$ (Fig.~\ref{fig:frozen}) of the gliders to measure the efficiency of energy extraction. We simulate gliders that execute random actions and gliders that have full information of the flow to obtain lower ($c_{\min}$) and upper bounds ($c_{\max}$) respectively of the climb rate. The efficiency of energy extraction is defined as
\begin{equation}
    \eta(c)=\frac{c-c_{\min}}{c_{\max}-c_{\min}}.
    \label{eq:effic}
\end{equation}
The efficiency estimates the value provided by global knowledge of the flow versus the local information acquired along the glider's path. Gliders with full information of this frozen flow show positive climb rates $c_{\max} = \SI{0.5}{m/s}$, implying that energy-positive flight is feasible if gliders have sufficient information and the flow has sufficiently long correlations. Gliders that use partial information---i.e., the measured wind velocities along their past trajectory---also show positive climb rates (Fig.~\ref{fig:frozen}, \ref{tab:climb_rate}) with $\eta = 0.77, 0.81$ for memory sizes $M = 20,50$ respectively (\ref{tab:efficiency}). The information is stored in the memory every $\Delta t=\SI{0.5}{s}$ and the memory duration is $t_m=M\Delta t$.

\subsection{Gliders in the dynamic wind field}
Next, we consider a dynamic wind field that has temporal correlations denoted by $\tau_c$. The frozen field considered above is obtained from the limit $\tau_c\to\infty$. To evaluate the performance, we measure the average climb rates of the gliders based on the altitude gained from \SI{200}{s} to \SI{500}{s}. We consider the frozen wind fields and dynamic wind fields where $\tau_c=\SI{16}{s}, \SI{32}{s}, \SI{64}{s} , \SI{128}{s}$. We then have full information gliders and partial information gliders with different memory sizes $M={4,10,20,30,40,50}$. Table \ref{tab:climb_rate} summarizes the details of the climb rates. Table \ref{tab:efficiency} presents the efficiencies of partial gliders in dynamic wind fields and frozen wind fields.

For smaller $\tau_c$, the reduced temporal correlations imply that past experience is less informative when picking out the optimal path via MCTS and thus the average climb rate decreases. Moreover, the intuition behind the spiraling bouts suggests that each bout will last at most $\sim \tau_c$, thereby reducing the time spent in energy-rich regions of the flow. When $\tau_c$ is sufficiently small, past information is not predictive of future flow configurations and the net energy gain should reduce to that of a random glider.

\begin{figure}
\centering
\includegraphics[width=12cm]{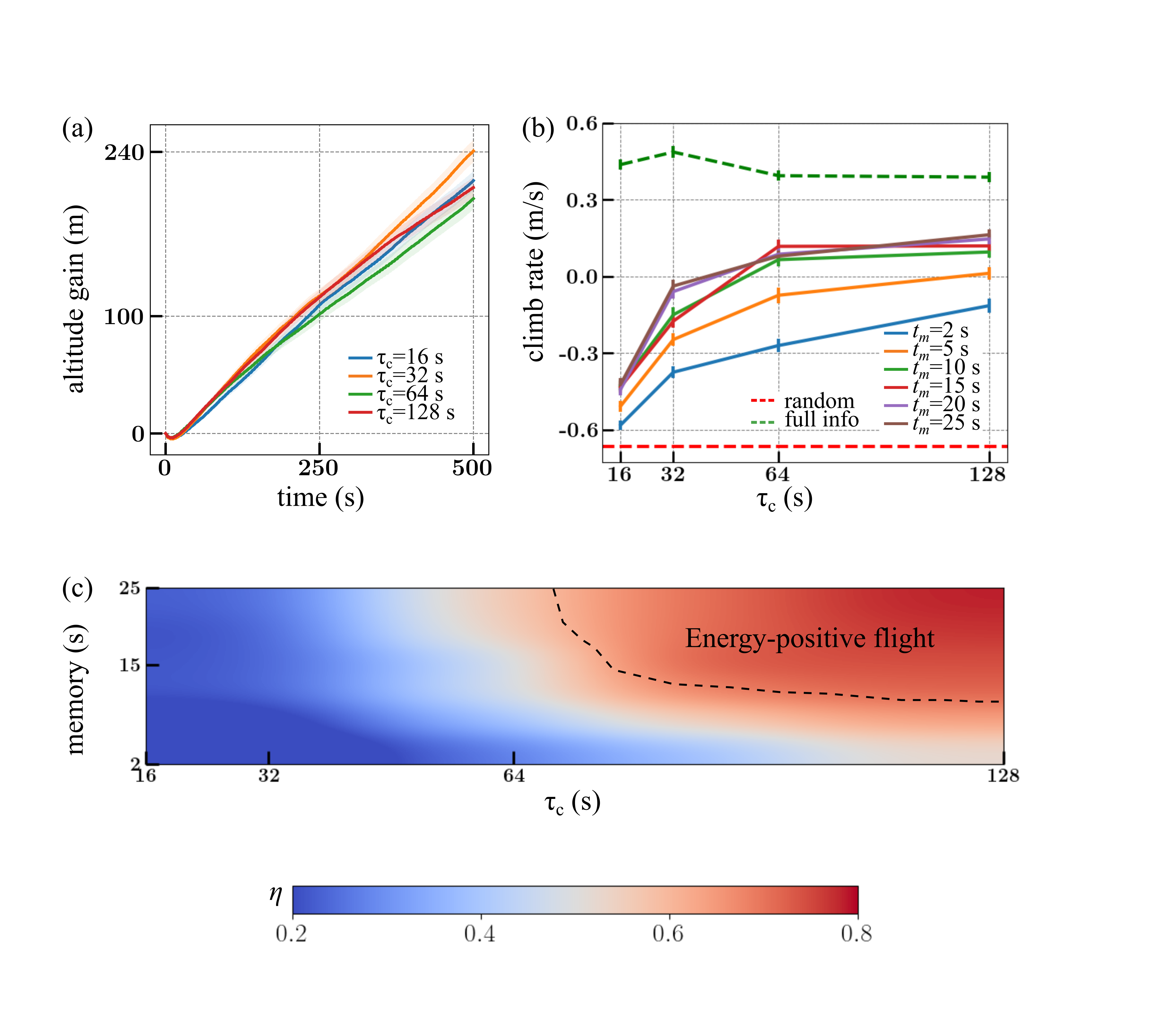}
\vspace{-1.5em}
\caption{(a) The average altitude gain of full information gliders in dynamic wind fields with $\tau_c=\SI{16}{s},\SI{32}{s},\SI{64}{s},\SI{128}{s}$. (b) The summary of climb rates with memory $t_m=\SI{5}{s},\SI{10}{s},\SI{15}{s},\SI{20}{s},\SI{25}{s}$ and $\tau_c=\SI{16}{s},\SI{32}{s},\SI{64}{s},\SI{128}{s}$. The climb rates of full information gliders are shown as the upper bound and the climb rates of random action gliders are shown as the lower bound. (c) Efficiency of energy extraction $\eta(c)$ as a function of $t_m$ and $\tau_c$. The energy-positive flight regime is where $c>0$.} \label{fig:dynamic_partial}
\end{figure}

Figure \ref{fig:dynamic_partial} shows that energy-positive flight is feasible for $\tau_c \gtrsim \SI{60}{s}$ for partial information gliders.
The net energy gain increases with $\tau_c$ and has weak dependence on memory size provided that $t_m > \SI{10}{s}$. The efficiency of the gliders increases monotonically with $\tau_c$ even though the climb rate when gliders have perfect information ($c_{\max}$) decreases with $\tau_c$, implying that the reduced climb rates in short time scale flows is due to the quicker degradation of acquired information rather than fewer sources of energy.

\begin{table}
\centering
\begin{tabular}{ccccccccc}
  \diagbox{$\tau_c (\text{s})$}{$t_m (\text{s})$} & \text{random} & 2 &  5 & 10    & 15    & 20    & 25    & \text{Full info.} \\
\hline
 16        & -0.67 & -0.581& -0.507 & -0.428 & -0.432 & -0.441 & -0.422 & 0.437  \\
 32 & -0.67 & -0.375& -0.247 & -0.150 & -0.175 & -0.060 & -0.038 &  0.486 \\
 64 & -0.67 & -0.270 & -0.074 & 0.065 &  0.117 &  0.086 &  0.079 &  0.394 \\
 128 & -0.67 & -0.115 & 0.012 & 0.095 & 0.119 &  0.146 & 0.162 & 0.388 \\
 frozen & -0.67 & -0.024 & 0.092 & 0.145 & 0.161 & 0.170 & 0.176 & 0.398 \\
\end{tabular}
\caption{\label{tab:climb_rate}Summary of climb rates (m/s) of partial information gliders and full information gliders in dynamic wind fields and frozen wind fields.}
\end{table}


\begin{table}
\centering
\begin{tabular}{ccccccc}
  \diagbox{$\tau_c (\text{s})$}{$t_m (\text{s})$}  & 2 &       5 & 10    & 15    & 20    & 25    \\
\hline
 16 &       7.74\% & 14.45\% & 21.66\% & 21.25\% & 20.44\% & 22.20\% \\
 32  &      25.30\% & 36.37\% & 44.79\% & 42.62\% & 52.65\% & 54.54\% \\
 64   &     37.35\% & 55.28\% & 68.27\% & 73.12\% & 70.24\% & 69.54\% \\
 128  &     52.32\% & 64.33\% & 72.24\% & 74.49\% & 77.06\% & 78.62\% \\
 frozen  &  60.38\% &71.25\% & 76.23\% & 77.81\% & 78.66\% & 79.16\% \\
\end{tabular}
\caption{\label{tab:efficiency}Summary of efficiencies (defined in Eq.~\eqref{eq:effic}) of partial information gliders in dynamic wind fields and frozen wind fields.}
\end{table}

\begin{figure}
    \centering
    \includegraphics[width=\textwidth]{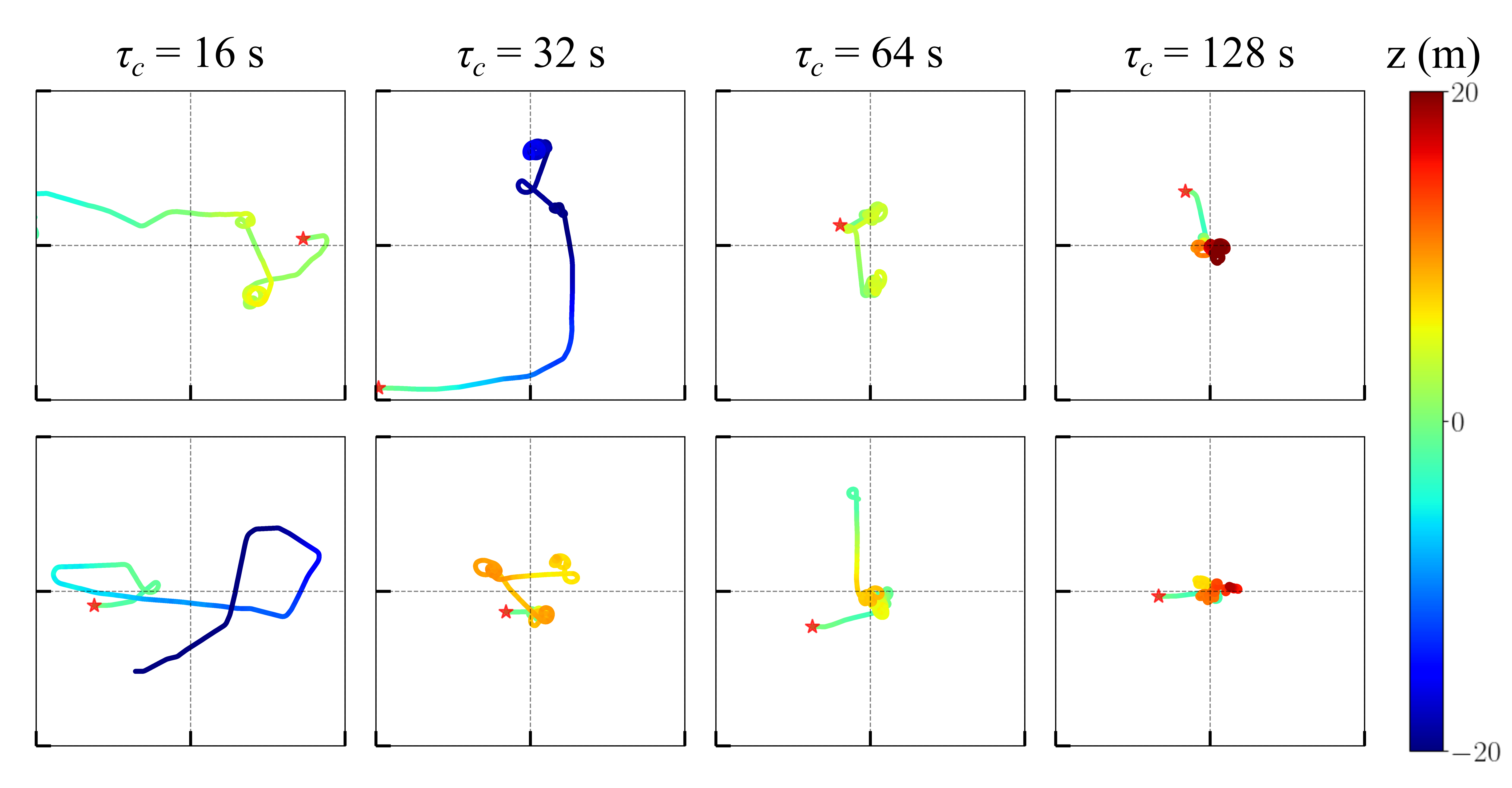}
    \caption{Examples of gliders trajectories in dynamic wind fields. Top view of gliders $\SI{100}{m}\times\SI{100}{m}$ $xy$-plane. The red star indicates the initial positions of the gliders. \label{fig:traj}}
\end{figure}

The partial information gliders demonstrate different exploration strategies under different field dynamics. As shown in Fig.~\ref{fig:traj}, for more fluctuating fields, the gliders discover and stop at different locations, while in more steady fields, the gliders behave more stably.

\section{Conclusion}
Soaring birds are believed to use relatively stable large-scale flow structures, such as thermals or shear layers, to reliably extract energy through efficient sampling and navigation. Our results show that  gliders employing a sampling-based planning algorithm can achieve energy-positive flight by exploiting transient fluctuations in a kinematic model of fluid turbulence. Rather than rely on local mechanical cues or optimized flight patterns, which have been shown to be useful for thermal soaring and dynamic soaring respectively, our algorithm uses a adaptive strategy where the recent history of measured wind velocities is used to predict and estimate the energy gained along future trajectories. Unlike those other modes of soaring, a memory of a few tens of seconds is necessary to achieve energy-positive flight, highlighting the importance of storing information to map out the local physical environment.

Most of the energy gained by gliders in our simulations is through localized updrafts, that is, the second term in the right hand side of Eq.~\eqref{eq:energyrate} rather than wind gradients. The trajectories of gliders reflect patterns of foraging, where efficient exploration of the flow through relatively straight paths is followed by energy extraction through localized spirals where significant updrafts are present. This picture suggests that the marginal benefit of a memory of a few tens of seconds is possibly a consequence of the correlation length scale of the flow (here $\approx \SI{100}{m}$): a glider that travels at \SI{10}{m/s} requires at least a memory of \SI{10}{s} to acquire two independent samples of the flow and guide its subsequent decisions.

Neutral conditions in the atmospheric boundary layer show correlation timescales of \SIrange{50}{100}{s} with typical velocity fluctuations of magnitude \SI{1}{m/s}~\citep{lenschow1986length}. Using simulated gliders that have glide-to-sink ratios (15:1) similar to soaring birds, our results suggest that energy-positive flight is indeed feasible at physically relevant timescales. Modern sailplanes travel at much faster speeds, allowing for more efficient exploration, and reach glide-to-sink ratios greater than 40:1~\citep{federal2004glider}. Sailplanes employing the algorithm devised here can potentially achieve energy-positive flight under stricter restrictions on atmospheric scales. The algorithm applies flows with arbitrary correlation structure, which include environments where updrafts have distinctive spatial arrangements (e.g., cloud streets in the atmospheric boundary layer~\citep{irving1973cloud}). An important caveat is that the predictive model relies on the second-order moments of the flow. An interesting open problem is whether non-Gaussian deviations in atmospheric turbulence~\citep{katul1994intermittency} significantly affect the algorithm's performance.

\backsection[Funding]{G.R.~was partially supported by the NSF-Simons Center for Mathematical \& Statistical Analysis of Biology at Harvard (award number \#1764269) and the Harvard Quantitative Biology Initiative. C.H.R.~was partially supported by the Applied Mathematics Program of the U.S.~DOE Office of Science Advanced Scientific Computing Research under Contract No.~DE-AC02-05CH11231.}

\backsection[Declaration of interests]{The authors report no conflict of interest.}

\backsection[Author ORCID]{D.\ He https://orcid.org/0009-0004-1100-9024; G.\ Reddy https://orcid.org/0000-0002-1276-9613; C.\ H.\ Rycroft, https://orcid.org/0000-0003-4677-6990}

\appendix

\section{Computation of the wind field}
\label{app:wind_num_imp}
As described in Sec.~\ref{sec:wind} the wind is modeled using an isotropic turbulence model comprised of $N\times N \times N$ Fourier modes (Eq.~\eqref{eq:dft2}) that evolve stochastically according to an Ornstein--Uhlenbeck process (Eq.~\eqref{eq:sde}). To visualize the wind field as in Fig.~\ref{fig:wind}, it is useful to perform a standard discrete Fourier transform at lattice locations using Eq.~\eqref{eq:dft}. To perform this, we make use of the FFTW library~\citep{frigo05}, and in particular the \texttt{c2r} transforms for transforming complex Fourier modes into a real field. Because of the restriction on the modes given by Eq.~\eqref{eq:mode_constr}, FFTW represents the modes in a $N\times N \times \lfloor \tfrac{N+1}{2} \rfloor$ array, where the other modes are implicitly defined. FFTW represents each complex number using a custom \texttt{fftw\_\,complex} data type, consisting of an array
of two double precision floating point numbers (occupying 16 bytes). We
therefore use this three-dimensional array as primary storage for the Fourier modes that describe the wind field.

Before performing a Fourier transform, FFTW uses its \texttt{fftw\_\,plan} data type to select the best algorithm for the particular data and memory layout. We use a custom \texttt{fftw\_\,plan} to perform the fast Fourier transforms for all three velocity components together, which due to vectorization provides some improvements in speed over doing three sequential computations. FFTW performs multithreaded computation and evaluates the complete three dimensional discrete Fourier transform in $O(N^3 \log N)$ time.

To evaluate the wind field at a single off-lattice glider position, we wrote a custom routine for computing Eq.~\eqref{eq:dft2}, which can be performed in $O(N^3)$ time. Because all the modes make independent computations to the sum in Eq.~\eqref{eq:dft2}, this computation can be efficiently multithreaded with OpenMP. Since this only has to compute a single velocity instead of a complete three-dimensional grid of velocities, it is considerably faster than the FFTW computation. In some cases, we consider a set of $G$ gliders within the same wind field, and we wrote an additional routine for computing $G$ velocities simultaneously. Since all of the $O(N^3)$ Fourier modes only need to be read from memory once, this routine is faster than performing $G$ individual wind evaluations.

Finally, we wrote a routine for updating the Fourier modes according to Eq.~\eqref{eq:sde}. We make use of the GNU Scientific Library (GSL) for computing the Gaussian random variables, with the Tausworthe random number generator~\citep{lecuyer96,lecuyer99}. This update can be multithreaded efficiently, with each thread using its own generator initialized with different seeds.

\begin{figure}
    \begin{center}
    \includegraphics[width=\textwidth]{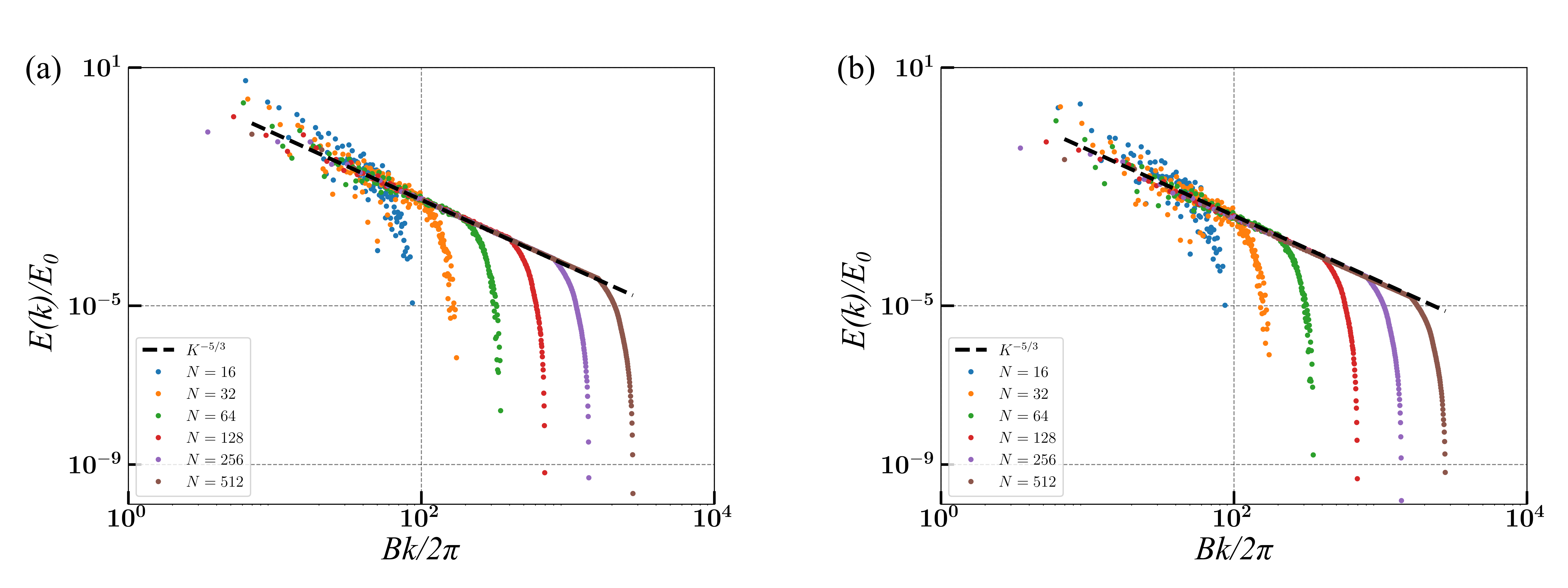}
    \end{center}
    \caption{The energy spectrum in simulations using different numbers of modes ($N=[16,32,64,128,256,512]$) captures Kolmogorov's $5/3$ law, where $E_0$ is a reference energy scale. (a) The initial wind field at $t=0$. (b) The wind field at $t=\SI{100}{s}$. \label{fig:law}}
\end{figure}

In Fig.~\ref{fig:law}, we show that the energy spectrum of the wind fields at the initial state and a later state $t=\SI{100}{s}$. Since the wind field is initialized at steady state, it satisfies Kolmogorov's power law that $E(k) \sim k^{-5/3}$. The later wind field also satisfies Kolmogorov's law, which demonstrates that that our stochastic differential equation in Eq.~\eqref{eq:sde} preserves the modes at equilibrium. As the number of modes $N$ is increased, Kolmogorov's law is extended to a larger range of $k$.

\section{Efficient Gaussian process regression}
\label{app:epgr}
As described in Sec.~\ref{sec:gpr}, a direct implementation of Eq.~\eqref{eq:posterior} would require computing $K(X,X)^{-1}$.
Inverting $K(X,X)$ each time, such as by using the LU decomposition~\citep{heath02}, will require $O(M^3)$ operations, which will become prohibitively expensive.  To improve performance, we therefore exploit that on each step $K(X,X)$ only changes slightly. Write $X=[x_1,x_2,\cdots,x_M]$. As gliders explore the field, the new memory replaces the old memory. When a new observation $x_{M+1}$ is added to $X$, the oldest observation $x_1$ is deleted, then we have $X'=[x_{M+1},x_2,\cdots,x_M]$. We can compute the new inverse of covariance function $K(X',X')^{-1}$ using $K(X,X)^{-1}$ efficiently by applying the Woodbury formula. Let $A=K(X,X), v=K(x_{M+1},X)$, and suppose $A^{-1}$ is known. Write
\[
K(X',X')^{-1}=(A+ev^\trans+ve^\trans)^{-1}.
\]
Where $e$ is a unit vector that $e_1=1$. According to the Woodbury formula
\[
(A+ev^\trans+ve^\trans)^{-1}=A^{-1}-A^{-1}U(C+VA^{-1}U)VA^{-1}
\]
where
\[
U=[e|v], \qquad V=U^\trans, \qquad C=\begin{bmatrix}
0 & 1 \\
1 & 0
\end{bmatrix}
\]
The update step requires $O(M^2)$ time, which is more efficient than computing the inverse of the matrix directly.

\begin{table}
\begin{center}
\begin{tabular}{lll}
\text{Symbol}  & Description & \text{Value}  \\
\hline
\text{Glider parameters} \\
\hline
$t$  & Time \\
$\br$ & Glider's position  \\
$\bu$ & Glider's ground velocity \\
$\bu_0$ & Glider's initial ground velocity & $\|\bu_0\|=\SI{10}{m/s}$ \\
$\bw$ & Wind velocity  \\
$w$   & Magnitude of wind velocity, $w=\|\bw\|$   \\
$\bar{w}$ & Mean of $w$ & $\SI{0.5}{m/s}$ \\
$\bv$ & Glider's air velocity, $\bv=\bu-\bw$ \\
$v$   & Glider's air speed, $v=\|\bv\|$ \\
$\rho$ & Air density \\
$S$   & Surface area of the wing \\
$c_L$ & Dimensionless lift coefficient    & 15 \\
$c_D$ & Dimensionless drag coefficient    & 1 \\
$L$   & Magnitude of the lift \\
$D$   & Magnitude of the drag \\
$m$   & Mass of glider \\
$g$   & Gravitational acceleration \\
$\mu$ & Glider's bank angle & $[-40^{\circ},40^{\circ}],\Delta \mu =\pm 10^{\circ}$\\
$\gamma$ & Angle of the glider's motion from horizontal   \\
$\psi$ & Glider's azimuth angle \\
$v_c$ & Speed scale                       & $\sqrt{2mg/\rho S}$\\
$t_c$ & Time scale                        & $v_c/g$ \\
$l_c$ & Length scale                      & $v_c^2/g$ \\
$E$   & Total energy of glider \\
$\varepsilon$ & Dimensionless total energy of glider \\
$G$  & Number of gliders in the same wind field & 20 \\
\hline
\text{Wind parameters} \\
\hline
$\bk$ & Wave number \\
$k$   & Magnitude of wave, $k=\|\bk\|$ \\
$\kmin$ & Smallest wave number & $\sqrt{3}\pi/B$ \\
$\kmax$ & Largest wave number & $\sqrt{3}\pi N/B$\\
$\Tilde{\bw}$ & Complex-valued amplitude of the wave   \\
$E(k)$ & Energy contained in waves \\
$N$   & Number of complex Fourier modes in one dimension \\
$\tau_c$ & Temporal correlations of the dynamic wind field& \SI{16}{s}, \SI{32}{s}, \SI{64}{s}, \SI{128}{s}\\
$\tau(k)$ & Time scale of the stochastic process  & \\
$K(r,t)$ & Correlation function of each wind component & \text{Eq.}~\eqref{eq:covariance_fun}\\
\end{tabular}
\caption{Parameters used throughout the paper for describing the glider and
wind field. \label{tab:param1}}
\end{center}
\end{table}

\begin{table}
\begin{center}
\begin{tabular}{lll}
\text{Symbol}  & Description & \text{Value}  \\
\hline
\multicolumn{2}{l}{\text{Gaussian process regression parameters}} \\
\hline
$M$ & Memory size & 4, 10, 20, 30, 40, 50 \\
$\Delta t$ & Time interval for one memory storage & \SI{0.5}{s} \\
$t_m$ & Memory duration &  \SI{2}{s}, \SI{5}{s}, \SI{10}{s}, \SI{15}{s}, \SI{20}{s}, \SI{25}{s}\\
\hline
\multicolumn{2}{l}{\text{Monte Carlo tree search parameters}} \\
\hline
$H_t$ & Glider's history at time $t$ \\
$V(H_t)$ & Maximum expected energy gained given $H_t$ \\
$d$ & Planning depth & 20 \\
\hline
\multicolumn{2}{l}{\text{Result parameters}} \\
\hline
$c$   & Climb rate of partial information glider (Table~\ref{tab:climb_rate}) & \\
$c_{min}$ & Sink rate &  $\SI{-0.67}{m/s}$ \\
$c_{max}$ & Climb rate of full information glider (Table~\ref{tab:climb_rate}) & \\
$\eta$    & Efficiency of partial information glider (Table~\ref{tab:efficiency}) & \\
\end{tabular}
\caption{Parameters used throughout the paper for describing the Gaussian process regression, Monte Carlo tree search, and results.\label{tab:param2}}
\end{center}
\end{table}

\bibliographystyle{jfm}
\bibliography{reference}
\end{document}